\documentclass[aps,twocolumn,a4paper,floatfix,superscriptaddress,prapplied,longbibliography]{revtex4-1}
\usepackage[toc,page]{appendix}
\usepackage{braket}
\usepackage{epsfig,graphicx,times}
\usepackage{amstext}
\usepackage{amsmath}
\usepackage{amssymb}
\usepackage{mathrsfs}
\usepackage{dcolumn}
\usepackage{bm}
\usepackage[tight]{subfigure}
\usepackage[colorlinks,linkcolor=blue,anchorcolor=blue,citecolor=blue,urlcolor=blue]{hyperref}
\usepackage{color}
\usepackage{siunitx}
\usepackage{appendix}
\usepackage{placeins}
\usepackage{textcomp}
\usepackage{bbold}
\usepackage{float}
\usepackage{verbatim}
\usepackage{minitoc}
\usepackage[dvipsnames]{xcolor}

\usepackage{soul}
\usepackage[framemethod=tikz]{mdframed}

\begin{document}

\title{Nonreciprocal enhancement of remote entanglement between nonidentical mechanical oscillators}

\author{Ya-Feng Jiao}
\affiliation{Key Laboratory of Low-Dimensional Quantum Structures and Quantum Control of Ministry of Education, \\
Department of Physics and Synergetic Innovation Center for Quantum Effects and Applications, \\ Hunan Normal University, Changsha 410081, China}
\affiliation{Laboratory of Chemical Biology $\&$ Traditional Chinese Medicine Research, \\
Ministry of Education College of Chemistry and Chemical Engineering, \\
Hunan Normal University, Changsha 410081, China}

\author{Jing-Xue Liu}
\affiliation{Key Laboratory of Low-Dimensional Quantum Structures and Quantum Control of Ministry of Education, \\
Department of Physics and Synergetic Innovation Center for Quantum Effects and Applications, \\
Hunan Normal University, Changsha 410081, China}

\author{Ying Li}
\affiliation{Key Laboratory of Low-Dimensional Quantum Structures and Quantum Control of Ministry of Education, \\
Department of Physics and Synergetic Innovation Center for Quantum Effects and Applications, \\
Hunan Normal University, Changsha 410081, China}

\author{Ronghua Yang}
\affiliation{Laboratory of Chemical Biology $\&$ Traditional Chinese Medicine Research, \\
Ministry of Education College of Chemistry and Chemical Engineering, \\ Hunan Normal University, Changsha 410081, China}

\author{Le-Man Kuang}\email{lmkuang@hunnu.edu.cn}
\affiliation{Key Laboratory of Low-Dimensional Quantum Structures and Quantum Control of Ministry of Education, \\
Department of Physics and Synergetic Innovation Center for Quantum Effects and Applications, \\ Hunan Normal University, Changsha 410081, China}
\affiliation{Synergetic Innovation Academy for Quantum Science and Technology, Zhengzhou University of Light Industry, Zhengzhou 450002, China}

\author{Hui Jing}\email{jinghui73@gmail.com}
\affiliation{Key Laboratory of Low-Dimensional Quantum Structures and Quantum Control of Ministry of Education, \\
Department of Physics and Synergetic Innovation Center for Quantum Effects and Applications, \\ Hunan Normal University, Changsha 410081, China}
\affiliation{Synergetic Innovation Academy for Quantum Science and Technology, Zhengzhou University of Light Industry, Zhengzhou 450002, China}

\date{\today}

\begin{abstract}
Entanglement between distant massive mechanical oscillators is of particular interest in quantum-enabled devices due to its potential applications in distributed quantum information processing. Here we propose how to achieve nonreciprocal remote entanglement between two spatially separated mechanical oscillators within a cascaded optomechanical configuration, where the two optomechanical resonators are indirectly coupled through a telecommunication fiber. We show that by selectively spinning the optomechanical resonators, one can break the time reversal symmetry of this compound system via Sagnac effect, and more excitingly, enhance the indirect couplings between the mechanical oscillators via the individual optimizations of light-motion interaction in each optomechanical resonator. This ability allows us to generate and manipulate nonreciprocal entanglement between distant mechanical oscillators, that is, the entanglement could be achieved only through driving the system from one specific input direction but not the other. Moreover, in the case of two frequency-mismatched mechanical oscillators, it is also found that the degree of the generated nonreciprocal entanglement is counterintuitively enhanced in comparison with its reciprocal counterparts, which are otherwise unattainable in static cascaded systems with a single-tone driving laser. Our work, which is well within the feasibility of current experimental capabilities, provides an enticing new opportunity to explore the nonclassical correlations between distant massive objects and facilitates a variety of emerging quantum technologies ranging from quantum information processing to quantum sensing.
\end{abstract}

\maketitle

\section{Introduction}

Entanglement, characterized by the abilities to explore strong nonclassical correlations between distinct quantum systems with spatial separation, is a peculiar feature of quantum physics~\cite{Horodecki2009RMP}. When sharing entanglement between different components in a composite quantum system, it can provide key quantum resources for a variety of nascent quantum technologies ranging from quantum information processing~\cite{Andersen2010LPR,Nielsen2010book,Flamini2018RPP} to quantum sensing~\cite{Degen2017RMP}, leading to the advances of quantum-enabled devices over its classical counterparts~\cite{Dowling2003PTRSA}. So far, although the concept of entanglement is still counterintuitive, it has been experimentally prepared, manipulated and utilized within a myriad of physical platforms involving both atomic scale particle and macroscopic scale objects~\cite{Raimond2001RMP,Pan2012RMP,Friis2019NRP}, e.g., photons~\cite{Yao2012NP,Wang2016PRL}, ions~\cite{Haffner2005Nature,Stute2012Nature}, atomic ensembles~\cite{Dai2016NP,Karg2020Science,Luo2022PRL}, and superconducting circuits~\cite{Campagne2018PRL,Kurpiers2018Nature}. Very recently, the generation of heralded entanglement between two $^{87}\textrm{Rb}$ atoms was even experimentally observed over fiber links with a length of $33\,\textrm{km}$~\cite{van2022Nature}, which is important for long-distance quantum communications~\cite{Cirac1997PRL}. Moreover, with the recent advances in the emerging field of cavity optomechanics (COM)~\cite{Aspelmeyer2014RMP,Xiong2015SCPMA,Barzanjeh2022NP}, the explorations of macroscopic quantum effects have been extended into a domain of massive mechanical oscillators~\cite{Connell2010Nature,Lai2018PRA,Lai2020PRA,Jahne2009PRA,Wollman2015Science,Xie2017PRA,Liao2016PRL}. For example, after subtracting the classical noise, nonclassical correlations were even demonstrated between light and $40\,\textrm{kg}$ mirrors at room temperature~\cite{Yu2020Nature}. In the past few years, thanks to the coherent light-motion interactions of COM systems~\cite{Verhagen2012Nature}, quantum entanglement has not only been intensively investigated theoretically and experimentally between electromagnetical field and vibrational modes~\cite{Vitali2007PRL,Palomaki2013Science,Ghobadi2014PRL,Riedinger2016Nature,Simon2018PRL,Li2022NP}, but also between propagating radiation fields~\cite{Tan2011PRA,Barzanjeh2019Nature,Chen2020NC} and between massive mechanical oscillators~\cite{Mancini2002PRL,Mancini2003EPJD,Genes2008NJP,Huang2009NJP,Tan2013PRA,Liao2014PRA,Yang2015PRA,Li2015NJP,Lai2022PRL}. Specially, by using multitone optical driving as an intermediary, the entangled state of distinct mechanical motions has been deterministically generated and directly observed with high fidelity in very recent experiments within optical or microwave COM systems~\cite{Riedinger2018Nature,Korppi2018Nature,Kotler2021Science,Mercier2021Science}.

\begin{figure*}[t]
\centering
% Requires \usepackage{graphicx}
\includegraphics[width=0.98\textwidth]{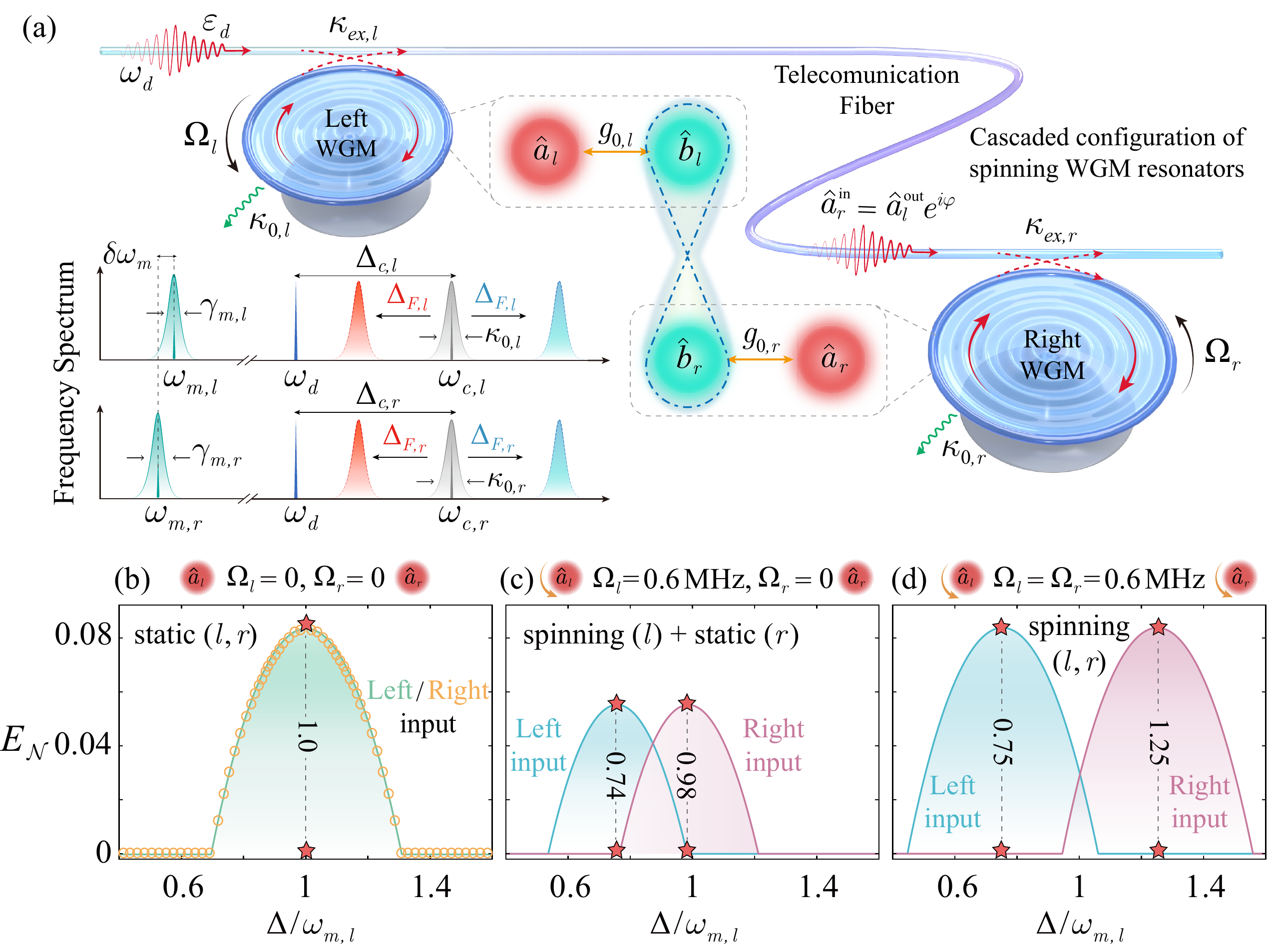}
\caption{\label{Fig_nonreciprocity}Nonreciprocal remote entanglement between two distant mechanical oscillators. (a) Schematic diagram and frequency spectrum of a unidirectional cascaded COM system composed of two spinning WGM resonators and a telecommunication fiber. Each WGM resonator supports two degenerate counterpropagating optical modes and a mechanical radial breathing mode, which are coupled through optical-radiation-pressure-mediated interactions. The two WGM resonators are assumed to have neither direct acoustic interaction nor near-field optical couplings, and their indirect coupling is enabled by the light field propagating in the fiber. By spinning the COM resonators along the CCW direction, we have $\Delta_{F,j}>0$ for the case of left input and $\Delta_{F,j}<0$ for the case of right input. By selectively spinning the WGM resonators and properly adjusting their spinning directions, the entanglement between the two distant mechanical oscillators becomes direction dependent, that is, it can be achieved when light is input from one specific direction but not the other. In the case of two identical mechanical oscillators, the logarithmic negativity $E_{\mathcal{N}}$ is plotted as a function of the scaled optical detuning $\Delta/\omega_{m,l}$ for different input directions, with angular velocities of the left- and right-hand-side WGM resonators (b) $\Omega_{l}=0,~\Omega_{r}=0$, (c) $\Omega_{l}=0.6\,\textrm{MHz},~\Omega_{r}=0$, and (d) $\Omega_{l}=0.6\,\textrm{MHz},~\Omega_{r}=0.6\,\textrm{MHz}$. In terms of the static case (i.e., $\Omega_{l}=\Omega_{r}=0$), $E_{\mathcal{N}}$ is present only within a small interval around the resonance $\Delta/\omega_{m,l}\simeq1$ and keeps reciprocal for the opposite input directions. In contrast, for the spinning case (i.e., $\Omega_{l}\neq0$ or $\Omega_{r}\neq0$), the resonance frequencies of the counterpropagating modes experience an opposite frequency shift due to the Sagnac effect, thereby leading to the peaks of $E_{\mathcal{N}}$ shifted for the opposite input directions.}
\end{figure*}

In parallel to these advances, nonreciprocal optical devices, serving as pivotal fundamental components in the fields of photonic quantum information processing~\cite{Flamini2018RPP}, have also attracted intense interest for their unique properties of unidirectional manipulation of light-matter interactions~\cite{Sounas2017NP,Lodahl2017Nature,Caloz2018PRApp}. In the past few decades, by breaking the time-reversal symmetry, several schemes have been proposed and experimentally demonstrated to fabricate nonreciprocal optical devices, which are generally based on magnetic materials~\cite{Shoji2014STAM}, atomic gases~\cite{Zhang2018NP,Yang2019PRL,Liang2020PRL}, optical or quantum nonlinearities~\cite{Dong2015NC,Kim2015NP,Xia2018prl,Hamann2018PRL,TangL2022PRL}, optomechanics~\cite{Manipatruni2009PRL,Shen2016NP,Bernier2017NC,Mercier2019prapp}, synthetic structures~\cite{Peng2014NP,Chang2014NP,Xu2020PRApp,Chen2021PRL}, and moving medium~\cite{Wang2013PRL,Lu2017PR,Maayani2018Nature}. As a crucial element for engineering the propagation of light, nonreciprocal optical devices have a wide range of applications, including invisible cloaking~\cite{Lin2011PRL}, unidirectional lasing~\cite{Bahari2017Science}, and suppression of backscattering~\cite{Kim2019Optica}. In recent years, nonreciprocal optical devices have also been extended to realize the one-way control of diverse physical phenomenon, such as optical chaos~\cite{Zhang2021PRA} or solitons~\cite{Li2021PRA}, phonon~\cite{Jiang2018PRApp} or magnon lasing~\cite{Huang2022OL}, and optomechanical high-order sidebands~\cite{LiWA2020PRA}. In view of advances in chiral quantum optics~\cite{Lodahl2017Nature}, nonreciprocal quantum devices have also been pursued very recently, which provide an indispensable tool for achieving, e.g., a single-photon isolator~\cite{Shen2011PRL,Dong2021SA,Ren2022LPR} or circulator~\cite{Scheucher2016Science,Tang2022PRL}, unidirectional flow of thermal noise~\cite{Barzanjeh2018PRL}, a one-way photon~\cite{Huang2018PRL,Li2019PR,Yang2019arxiv} or magnon blockade~\cite{Wang2022SCPMA}, and nonreciprocal entanglement~\cite{Jiao2020PRL,Orr2022arxiv}. In particular, we note that the nonreciprocal entanglement between optical field and mechanical oscillators was proposed very recently by using a single spinning microsphere resonator, which reveals its counterintuitive property as being robust against random losses in a chosen direction~\cite{Jiao2020PRL}.

In this paper, we propose how to achieve nonreciprocal remote entanglement between two distant massive mechanical oscillators in a cascaded COM system consisting of two spinning whispering-gallery-mode (WGM) resonators and a telecommunication fiber, unveiling its unique properties that are otherwise unattainable in its reciprocal counterparts. Specifically, by selectively spinning the WGM resonators and properly adjusting their spinning directions, we present an effective method to generate and manipulate nonreciprocal entanglement between two distant mechanical oscillators. More excitingly, we find that in the case of nonidentical mechanical oscillators, especially of those with frequency mismatch, the achievement of such nonreciprocal entanglement provides an enticing new opportunity to enhance its entanglement degree. The Sagnac effect induced by spinning devices plays a key role in this process. It allows us to break the time-reversal symmetry of this compound system, and also to enhance the optical-radiation-pressure-mediated couplings between mechanical oscillators via the individual optimization of light-motion interaction in each WGM resonator. These results are established with parameters that are well within the feasibilities of current experimental capabilities. As such, we anticipate that our work could serve as a useful tool to explore the nonclassical natures of composite systems involving massive mechanical objects~\cite{Mancini2002PRL,Mancini2003EPJD,Genes2008NJP,Huang2009NJP,Tan2013PRA,Liao2014PRA,Yang2015PRA,
Li2015NJP,Lai2022PRL,Riedinger2018Nature,Korppi2018Nature,Kotler2021Science,Mercier2021Science}, open up promising ways to engineer various quantum effects with diverse nonreciprocal devices~\cite{Sounas2017NP,Lodahl2017Nature,Caloz2018PRApp}, and provide crucial quantum resources to a large number of emerging quantum technologies ranging from quantum information processing~\cite{Andersen2010LPR,Nielsen2010book,Flamini2018RPP} to quantum sensing~\cite{Degen2017RMP}. For example, a network of nonreciprocal entangled sensors would be useful for distributed quantum sensing or other quantum network applications~\cite{Cirac1997PRL}.

This paper is structured as follows. In Sec.\,\ref{SecII}, we introduce the theoretical model of the cascaded spinning COM system and derive the linearized dynamics and the covariance matrix of this system. In Sec.\,\ref{SecIII}, the generation and manipulation of nonreciprocal entanglement between distant mechanical oscillators are explicitly studied. In particular, the capability of our proposed scheme to directionally enhance the entanglement between nonidentical mechanical oscillators is revealed. In Sec.\,\ref{SecIV}, a summary is given.

\section{Theoretical model and dynamics of a cascaded spinning COM system}\label{SecII}

As studied in detail in Refs.~\cite{Mancini2002PRL,Mancini2003EPJD,Genes2008NJP,Huang2009NJP,Tan2013PRA,Liao2014PRA,Yang2015PRA,
Li2015NJP,Lai2022PRL,Riedinger2018Nature,Korppi2018Nature,Kotler2021Science,Mercier2021Science}, for two spatially separated mechanical objects, the macroscopic entanglement between them can be generated with the help of an optical-radiation-pressure-mediated interaction. Meanwhile, it is shown that the emergence of such macroscopic entanglement by using a single-tone optical driving occurs only within a small bandwidth around the resonant mechanical parameter region~\cite{Mancini2003EPJD,Genes2008NJP}, i.e., indicating that this single-tone driving scheme requires each mechanical oscillator with identical COM coupling rate, resonance frequency, and damping rate. In terms of creating and detecting entanglement between nonidentical mechanical oscillators, especially between those with mismatched fundamental frequencies, an effective approach is to apply a strategy that involves a multitone optical driving process~\cite{Li2015NJP}, in which each tone of the multiplexing laser is assumed to be associated with a specific mechanical oscillator. Important experimental milestone researches toward this goal have been reached very recently by using pulsed multitone driving as an intermediary in optical or microwave COM systems~\cite{Riedinger2018Nature,Korppi2018Nature,Kotler2021Science,Mercier2021Science}. However, the main constraints on efficiencies of this multitone driving scheme, in comparison with its counterpart performed by single-tone driving, is that the multitone driving scheme works optimally under the rotating wave approximation (RWA) condition~\cite{Li2015NJP}, and thus it requires the fundamental mechanical frequencies $\omega_{m}$ and the corresponding frequency difference $\delta\omega_{m}$ to be much larger than the cavity bandwidth $\kappa$, i.e., $\omega_{m},\,\delta\omega_{m}\gg\kappa$. Here the frequency mismatch $\delta\omega_{m}$ is defined as $\delta\omega_{m}/\omega_{m,l}=1-\chi$, with a mechanical frequency ratio $\chi=\omega_{m,r}/\omega_{m,l}$. For current experimental techniques with optical cavities, it still remains a tough challenge to satisfy the above RWA condition for small values of $\delta\omega_{m}$, which requires ultrahigh quality factors of cavity modes. Therefore, it is necessary to establish an effective method to create and manipulate remote entanglement between nonidentical distant mechanical oscillators.

In this paper, we propose how to achieve nonreciprocal remote entanglement between two distant massive mechanical oscillators, unveiling its unique properties that are otherwise unattainable in reciprocal systems. Specifically, as shown in Fig.\,\ref{Fig_nonreciprocity}(a), we consider a unidirectional cascaded COM system consisting of two spatially distant WGM resonators and a telecommunication fiber. WGM resonators, such as microdisks, microspheres, and microtoroids, usually support two degenerate counterpropagating optical modes, whose typical quality factors (Q) can achieve $Q\sim10^5-10^9$~\cite{Xiao2020book}. By design, each WGM resonator can also support a mechanical breathing mode, resulting from its radial vibration induced by optical radiation pressure. Here we denote the fundamental frequencies of the optical and mechanical modes by $\omega_{c,j} $ and $ \omega_{m,j} $, respectively, where the index $ j\!=\!l,r $ refers to the WGM resonator placed on the left- or right-hand side. In addition, the two WGM resonators are also assumed to have neither direct acoustic interaction nor near-field optical couplings, whereas the localized optical and mechanical modes of WGM resonators are coupled with each other due to the optical-radiation-pressure interactions. Furthermore, we note that the input optical fields in the fiber can be evanescently coupled into and out of the $j$-hand-side WGM resonator (with a coupling rate $\kappa_{\textit{ex},j}$), and propagate between the two WGM resonators unidirectionally, thereby leading to an indirect intercavity unidirectional coupling between the optical modes~\cite{Xiao2010PRA,Li2016OE}. We remark that by utilizing such cascaded configurations, frequency locking or synchronization of distant nanomechanical oscillators has been experimentally demonstrated very recently~\cite{Santos2017PRL,Zhang2015PRL,Li2022PRL}.

Based on this compound system, an effective way for achieving optical nonreciprocity is to break the time-reversal symmetry of this system through selectively spinning the WGM resonators~\cite{Qin2020APE}. In particular, nonreciprocal transmission with $ 99.6 \% $ isolation was experimentally observed for light within a spinning WGM sphere very recently~\cite{Maayani2018Nature}, without the facility of any magnetic materials, optical nonlinearities, and complex spatial structures. In this experiment, for achieving continued rotation, the WGM resonator is mounted on a turbine and spins along its axis. By properly positioning a telecommunication fiber near such a spinning resonator, stable fiber-resonator coupling is established via the ``self-adjustment" aerodynamic process, which thus enables the evanescent coupling of light into and out of the spinning resonators. Also, for such spinning devices, counterpropagating light generally experiences different effective optical path lengths as it travels for a whole enclosed loop, thus leading to an irreversible refractive index for the clockwise (CW) and counterclockwise (CCW) modes~\cite{Maayani2018Nature}, that is, $ n_{\pm}\!=\!n[1\pm nR\Omega(n^{-2}-1)/c] $, where $ n $ and $ \Omega $ are the refractive index and the angular velocity of a WGM resonator with radius $ R $, respectively, and $ c $ is the speed of light in vacuum. As a result, the degeneracy of the CW and CCW optical modes is correspondingly lifted by a Fizeau frequency shift $\Delta_{F}$, resulting from the well-known Sagnac effect. Moreover, for the case of light input from the left (right) direction, we have $\Delta_{F,j}>0$ ($\Delta_{F,j}<0$) for the $j$-hand-side WGM resonator spinning along the CCW direction, and the corresponding effective resonance frequencies of the counterpropagating modes are given by
\begin{equation}
\centering
\omega_{j,s}\!\equiv\!\omega_{c,j}\pm|\Delta_{F,j}|,
\end{equation}
with $s=\circlearrowright,\circlearrowleft$ indexing the CW or CCW optical modes, and
\begin{equation}
\centering
\Delta_{F,j}=\pm\Omega_{j}\dfrac{n_{j}R_{j}\omega_{c,j}}{c}\left(1-\dfrac{1}{n_{j}^{2}}-\dfrac{\lambda_{j}}{n_{j}}\dfrac{dn_{j}}{d\lambda_{j}}\right), \label{Sagfs}
\end{equation}
where $ \lambda_{j}=c/\omega_{c,j} $ is the wavelength of light in vacuum. The dispersion term $ dn_{j}/d\lambda_{j} $, characterizing the relativistic origin of the Sagnac effect, is relatively small in typical materials (reaching up to about $ 1\% $)~\cite{Maayani2018Nature,Malykin2000PU}.

In a rotating frame with respect to $\hat{H}_{d}=\hbar\omega_{d}(\hat{a}_{l}^{\dagger}\hat{a}_{l}+\hat{a}_{r}^{\dagger}\hat{a}_{r})$, the Hamiltonian of this compound system, with each WGM resonator spinning along the CCW direction and the optical driving field input from the left-hand side, is given by
\begin{align}
\hat{H}=&\hat{H}_{0}+\hat{H}_{\textrm{int}}+\hat{H}_{\textrm{dr}},\notag\\
\hat{H}_{0}=&\hbar\Delta_{l,\circlearrowright}\hat{a}_{l}^{\dagger}\hat{a}_{l}+\hbar\Delta_{r,\circlearrowright}\hat{a}_{r}^{\dagger}\hat{a}_{r}
+\hbar\omega_{m,l}\hat{b}_{l}^{\dagger}\hat{b}_{l}+\hbar\omega_{m,r}\hat{b}_{r}^{\dagger}\hat{b}_{r},\notag\\
\hat{H}_{\textrm{int}}=&-\hbar g_{0,l}\hat{a}_{l}^{\dagger}\hat{a}_{l}(\hat{b}_{l}^{\dagger}+\hat{b}_{l})
-\hbar g_{0,r}\hat{a}_{r}^{\dagger}\hat{a}_{r}(\hat{b}_{r}^{\dagger}+\hat{b}_{r}),
\notag\\
\hat{H}_{\textrm{dr}}=&i\hbar\sqrt{\kappa_{\textit{ex},l}}\varepsilon_{d}(\hat{a}_{l}^{\dagger}-\hat{a}_{l}),
\label{Hamiltonian}
\end{align}
where $\hat{a}_{j}$ ($\hat{a}_{j}^{\dagger}$) and $\hat{b}_{j}$ ($\hat{b}_{j}^{\dagger}$) are respectively the optical and mechanical annihilation (creation) operators, and $\Delta_{j,\circlearrowright}=\omega_{j,\circlearrowright}-\omega_{d}$ denotes the optical detuning between the spinning resonator and the driving field. The corresponding frame rotating with driving frequency $\omega_{d}$ is implemented by applying a unitary transformation, i.e., $\hat{H} = \hat{U}\hat{H}(t)\hat{U}^{\dagger}-i\hbar\hat{U}\partial\hat{U}^{\dagger}/\partial t$, with $\hat{U}=\exp[i\hat{H}_{d}t/\hbar]$ (see, e.g., Refs.\,\cite{Aspelmeyer2014RMP} for more details). The parameter $g_{0,j}=(\omega_{c,j}/R_{j})\sqrt{\hbar/m_{j}\omega_{m,j}}$ denotes the single-photon COM coupling strength of the $j$-hand-side WGM resonator, with an effective mass $m_{j}$. The field amplitude of the driving laser is given by $|\varepsilon_{d}|=\sqrt{P_{d}/\hbar\omega_{d}} $, and $P_{d}$ is referred to as the input laser power of the driving field. Note that for studying the case where the driving field is input from the right-hand side, one should alter the index of the optical detuning, i.e., $\Delta_{j,\circlearrowright}\rightarrow\Delta_{j,\circlearrowleft}$, and change the Hamiltonian of the driving field to $\hat{H}_{\textrm{dr}}=i\hbar\sqrt{\kappa_{\textit{ex},r}}\varepsilon_{d}(\hat{a}_{r}^{\dagger}-\hat{a}_{r})$.

In the following discussions, as a specific example, we focus on the left-hand-side driving case and show how to explore the dynamics of this compound system. By introducing dissipations and input noise, the equations of motion for arbitrary optical and mechanical operators can be obtained through employing the quantum Langevin equations (QLEs), which are given by
\begin{align}
\dot{\hat{a}}_{l}=&-(i\Delta_{l,\circlearrowright}+\dfrac{\kappa_{0,l}+\kappa_{\textit{ex},l}}{2})\hat{a}_{l}
+ig_{0,l}\hat{a}_{l}(\hat{b}_{l}^{\dagger}+\hat{b}_{l}) \notag\\
&+\sqrt{\kappa_{\textit{ex},l}}\varepsilon_{d}
+\sqrt{\kappa_{0,l}}\hat{a}_{0,l}^{\textrm{in}}
+\sqrt{\kappa_{\textit{ex},l}}\hat{a}_{\textit{ex},l}^{\textrm{in}},\notag\\
%%%%%%%%%%%%%%%%%%%%%%%%%%%%%%%%%%%%%%%%%%%%%%%%%%%%%%%%%%%%%%%%%%%%%%%%%%%%%%%%%%%%%%%%%%%%%%%
\dot{\hat{a}}_{r}=&-(i\Delta_{r,\circlearrowright}+\dfrac{\kappa_{0,r}+\kappa_{\textit{ex},r}}{2})\hat{a}_{r}
+ig_{0,r}\hat{a}_{r}(\hat{b}_{r}^{\dagger}+\hat{b}_{r}) \notag\\
&+\sqrt{\eta\kappa_{\textit{ex},r}}\hat{a}_{f,r}^{\textrm{in}}
+\sqrt{\kappa_{0,r}}\hat{a}_{0,r}^{\textrm{in}}
+\sqrt{\kappa_{\textit{ex},r}}\hat{a}_{\textit{ex},r}^{\textrm{in}},\notag \\
%%%%%%%%%%%%%%%%%%%%%%%%%%%%%%%%%%%%%%%%%%%%%%%%%%%%%%%%%%%%%%%%%%%%%%%%%%%%%%%%%%%%%%%%%%%%%%%
\dot{\hat{b}}_{l}=&-(i\omega_{m,l}+\dfrac{\gamma_{m,l}}{2})\hat{b}_{l}+ig_{0,l}\hat{a}_{l}^{\dagger}\hat{a}_{l}+\sqrt{\gamma_{m,l}}\hat{b}_{l}^{\textrm{in}},\notag \\
\dot{\hat{b}}_{r}=&-(i\omega_{m,r}+\dfrac{\gamma_{m,r}}{2})\hat{b}_{r}+ig_{0,r}\hat{a}_{r}^{\dagger}\hat{a}_{r}+\sqrt{\gamma_{m,r}}\hat{b}_{r}^{\textrm{in}},
\label{qles}
\end{align}
where $\kappa_{0,j}$ and $\gamma_{m,j}$ are the intrinsic decay rates of the optical and mechanical modes, respectively. $\hat{a}_{0,j}^{\textrm{in}}$ and $\hat{a}_{\textit{ex},j}^{\textrm{in}}$ describe two distinct types of Gaussian noise caused by the zero-point fluctuations of the optical field from vacuum and the fiber mode, and $\hat{b}_{j}^{\textrm{in}}$ denotes the input vacuum noise operators for the mechanical modes. These input vacuum noise operators have zero mean values and are characterized by the correlation functions~\cite{Zoller2000book}
\begin{align}
\left\langle\hat{a}_{0,j}^{\textrm{in}}(t)\hat{a}_{0,j^{\prime}}^{\textrm{in},\dagger}(t^{\prime})\right\rangle=&\delta_{jj^{\prime}}\delta(t-t^{\prime}), \notag\\
\left\langle\hat{a}_{\textit{ex},j}^{\textrm{in}}(t)\hat{a}_{\textit{ex},j^{\prime}}^{\textrm{in},\dagger}(t^{\prime})\right\rangle=&\delta_{jj^{\prime}}\delta(t-t^{\prime}), \notag\\
\left\langle\hat{b}_{j}^{\textrm{in}}(t)\hat{b}_{j^{\prime}}^{\textrm{in},\dagger}(t^{\prime})\right\rangle=&(\bar{n}_{m,j}+1)\delta_{jj^{\prime}}\delta(t-t^{\prime}),
\notag\\
\left\langle\hat{b}_{j}^{\textrm{in},\dagger}(t)\hat{b}_{j^{\prime}}^{\textrm{in}}(t^{\prime})\right\rangle=&\bar{n}_{m,j}\delta_{jj^{\prime}}\delta(t-t^{\prime}),
\end{align}
where $\bar{n}_{m,j}=[\textrm{exp}(\hbar\omega_{m,j}/\textrm{k}_{\textrm{B}}\textrm{T})-1]^{-1}$ is the thermal mean occupation number of the mechanical mode, $\textrm{k}_{\textrm{B}}$ is the Boltzmann constant, and $\textrm{T}$ is the bath temperature of the mechanical mode. Here we have assumed that the cavity modes are in ordinary zero-temperature environments, whereas the mechanical modes are surrounded by thermal reservoirs. In addition, the operator $\hat{a}_{f,r}^{\textrm{in}}$ denotes the coherent input optical field of the right-hand-side WGM resonator. In our cascaded COM system, for the optical driving field input from the left direction, it first excites the CW mode of the left-hand-side WGM resonator, then couples out of this resonator to the fiber, and, subsequently, the corresponding output field $\hat{a}_{l}^{\textrm{out}}$ is injected into the right-hand-side WGM resonator after propagating for a distance in the fiber, which excites the CW mode as well. Based on such properties of the cascaded systems and the input-output relations, we have~\cite{Xiao2010PRA,Li2016OE,Santos2017PRL,Zhang2015PRL,Li2022PRL}
\begin{align}
\hat{a}_{l}^{\textrm{out}}=&\varepsilon_{d}-\sqrt{\kappa_{\textit{ex},l}}\hat{a}_{l},
\notag \\
\hat{a}_{f,r}^{\textrm{in}}=&\hat{a}_{l}^{\textrm{out}}e^{i\varphi},
\end{align}
where the additional phase $\varphi$ describes the accumulated phase delay resulting from light propagation between the two spinning WGM resonators, and it is given by
\begin{align}
\varphi\equiv\omega_{d}\tau=2\pi\dfrac{nL}{\lambda},
\end{align}
where $\tau$ denotes the propagating time and $L$ is the length of the fiber between the coupling regions of the two WGM resonators. Moreover, $\eta\in[0,1]$ represents the power transmission coefficient of the propagating field in the fiber, which accounts for the imperfect couplings between the two spinning WGM resonators. For $\eta=1$, it indicates the case of a lossless unidirectional coupling, and in contrast, $\eta=0$ corresponds to the situation where the two spinning WGM resonators are independent without any optical coupling.

QLEs (\ref{qles}) involve the radiation-pressure-induced nonlinear COM interactions between optical fields and mechanical oscillators, and thus are difficult to be directly solved. In order to find solutions to such a set of nonlinear dynamical equations, one can expand each operator as a sum of its steady-state mean value and a small quantum fluctuation around it under the condition of strong optical driving, i.e.,
\begin{align}
\hat{a}_{j}=\alpha_{j}+\delta\hat{a}_{j}, \notag\\
\hat{b}_{j}=\beta_{j}+\delta\hat{b}_{j}.
\end{align}
By substituting these assumptions into QLEs.\,(\ref{qles}), we obtain the first-order inhomogeneous differential equations for steady-state mean values
\begin{align}
\dot{\alpha}_{l}=&-(i\tilde{\Delta}_{l,\circlearrowright}+\dfrac{\kappa_{0,l}+\kappa_{\textit{ex},l}}{2})\alpha_{l}
+\sqrt{\kappa_{\textit{ex},l}}\varepsilon_{d},\notag\\
%%%%%%%%%%%%%%%%%%%%%%%%%%%%%%%%%%%%%%%%%%%%%%%%%%%%%%%%%%%%%%%%%%%%%%%%%%%%%%%%%%%%%%%%%%%%%%%
\dot{\alpha}_{r}=&-(i\tilde{\Delta}_{r,\circlearrowright}+\dfrac{\kappa_{0,r}+\kappa_{\textit{ex},r}}{2})\alpha_{r}
-\sqrt{\eta\kappa_{\textit{ex},l}\kappa_{\textit{ex},r}}e^{i\varphi}\alpha_{l}
\notag\\
&+\sqrt{\eta\kappa_{\textit{ex},r}}e^{i\varphi}\varepsilon_{d},
\notag\\
%%%%%%%%%%%%%%%%%%%%%%%%%%%%%%%%%%%%%%%%%%%%%%%%%%%%%%%%%%%%%%%%%%%%%%%%%%%%%%%%%%%%%%%%%%%%%%%
\dot{\beta}_{l}=&-(i\omega_{m,l}+\dfrac{\gamma_{m,l}}{2})\beta_{l}+ig_{0,l}|\alpha_{l}|^{2},\notag \\
\dot{\beta}_{r}=&-(i\omega_{m,r}+\dfrac{\gamma_{m,r}}{2})\beta_{r}+ig_{0,r}|\alpha_{r}|^{2},
\label{steady-state}
\end{align}
and the corresponding linearized QLEs for quantum fluctuations
\begin{align}
\delta\dot{\hat{a}}_{l}=&-(i\tilde{\Delta}_{l,\circlearrowright}+\dfrac{\kappa_{0,l}+\kappa_{\textit{ex},l}}{2})\delta\hat{a}_{l}
+ig_{0,l}\alpha_{l}(\delta\hat{b}_{l}^{\dagger}+\delta\hat{b}_{l}) \notag\\
&+\sqrt{\kappa_{0,l}}\hat{a}_{0,l}^{\textrm{in}}
+\sqrt{\kappa_{\textit{ex},l}}\hat{a}_{\textit{ex},l}^{\textrm{in}},\notag\\
%%%%%%%%%%%%%%%%%%%%%%%%%%%%%%%%%%%%%%%%%%%%%%%%%%%%%%%%%%%%%%%%%%%%%%%%%%%%%%%%%%%%%%%%%%%%%%%
\delta\dot{\hat{a}}_{r}=&-(i\tilde{\Delta}_{r,\circlearrowright}+\dfrac{\kappa_{0,r}+\kappa_{\textit{ex},r}}{2})\delta\hat{a}_{r}
-\sqrt{\eta\kappa_{\textit{ex},l}\kappa_{\textit{ex},r}}e^{i\varphi}\delta\hat{a}_{l} \notag\\
&+ig_{0,r}\alpha_{r}(\delta\hat{b}_{r}^{\dagger}+\delta\hat{b}_{r})
+\sqrt{\kappa_{0,r}}\hat{a}_{0,r}^{\textrm{in}}
+\sqrt{\kappa_{\textit{ex},r}}\hat{a}_{\textit{ex},r}^{\textrm{in}},\notag \\
%%%%%%%%%%%%%%%%%%%%%%%%%%%%%%%%%%%%%%%%%%%%%%%%%%%%%%%%%%%%%%%%%%%%%%%%%%%%%%%%%%%%%%%%%%%%%%%
\delta\dot{\hat{b}}_{l}=&-(i\omega_{m,l}+\dfrac{\gamma_{m,l}}{2})\delta\hat{b}_{l}+ig_{0,l}(\alpha_{l}\delta\hat{a}_{l}^{\dagger}+\alpha_{l}^{\ast}\delta\hat{a}_{l}) \notag \\
&+\sqrt{\gamma_{m,l}}\hat{b}_{l}^{\textrm{in}},\notag \\
\delta\dot{\hat{b}}_{r}=&-(i\omega_{m,r}+\dfrac{\gamma_{m,r}}{2})\delta\hat{b}_{r}+ig_{0,r}(\alpha_{r}\delta\hat{a}_{r}^{\dagger}+\alpha_{r}^{\ast}\delta\hat{a}_{r}) \notag \\
&+\sqrt{\gamma_{m,r}}\hat{b}_{r}^{\textrm{in}},
\label{qfluctuation}
\end{align}
where $\tilde{\Delta}_{j,\circlearrowright}=\Delta_{c,j}+|\Delta_{F,j}|$, and $\Delta_{c,j}=(\omega_{c,j}-\omega_{d})-g_{0,j}(\beta_{j}^{\ast}+\beta_{j})$ is an effective optical detuning caused by a radiation-pressure-mediated phase shift. By setting all derivatives in Eq.\,(\ref{steady-state}) to be zero, the steady-state mean values of the optical and mechanical modes can be obtained as
\begin{align}
\alpha_{l}=&\dfrac{\sqrt{\kappa_{\textit{ex},l}}\varepsilon_{d}}{i\tilde{\Delta}_{l,\circlearrowright}+(\kappa_{0,l}+\kappa_{\textit{ex},l})/2},\notag \\
\alpha_{r}=&\dfrac{\sqrt{\eta\kappa_{\textit{ex},r}}e^{i\varphi}(\varepsilon_{d}-\sqrt{\kappa_{\textit{ex},l}}\alpha_{l})}{i\tilde{\Delta}_{r,\circlearrowright}+(\kappa_{0,r}+\kappa_{\textit{ex},r})/2}, \notag\\
\beta_{j}=&\dfrac{ig_{0,j}|\alpha_{j}|^{2}}{i\omega_{m,j}+\gamma_{m,j}/2}.
\label{SSmeanvalue}
\end{align}
Equation\,(\ref{SSmeanvalue}) indicates that the intracavity photon number, $N_{j}\equiv|\alpha_{j}|^{2}$, of the left-hand-side WGM resonator is merely related to the pump strength of the input driving field and, simultaneously, it also dominates the intracavity photon number of the right-hand-side WGM resonator. In addition, it is clearly seen that although the field amplitude of the right-hand-side COM resonator is associated with the propagating phase delay $\varphi$, the intracavity photon number of such a WGM resonator is independent of it, implying that the effective COM coupling strength of the right-hand-side COM resonator is identical for arbitrary propagating distance in lossless unidirectional cascaded COM systems.

By defining the optical and mechanical quadrature fluctuation operators as
\begin{align}
\nonumber
\hat{X}_{j}&=\dfrac{1}{\sqrt{2}}(\delta\hat{a}_{j}^{\dagger}
+\delta\hat{a}_{j}),
~~~\hat{Y}_{j}=\dfrac{i}{\sqrt{2}}
(\delta\hat{a}_{j}^{\dagger}-\delta\hat{a}_{j}), \notag\\
\hat{q}_{j}&=\dfrac{1}{\sqrt{2}}(\delta\hat{b}_{j}^{\dagger}+\delta\hat{b}_{j}),
~~~\hat{p}_{j}=\dfrac{i}{\sqrt{2}}(\delta\hat{b}_{j}^{\dagger}-\delta\hat{b}_{j}),
\end{align}
and the associated Hermitian input noise operators as
\begin{align}
\hat{x}_{j}^{\textrm{in}}&=\dfrac{1}{\sqrt{2}}
(\hat{a}_{j}^{\textrm{in}\dagger}+\hat{a}_{j}^{\textrm{in}}),~~~
\hat{y}_{j}^{\textrm{in}}=\dfrac{i}{\sqrt{2}}(\hat{a}_{j}^{\textrm{in}\dagger}
-\hat{a}_{j}^{\textrm{in}}),\notag\\
\hat{q}_{j}^{\textrm{in}}&=\dfrac{1}{\sqrt{2}}
(\hat{b}_{j}^{\textrm{in}\dagger}+\hat{b}_{j}^{\textrm{in}}),~~~
\hat{p}_{j}^{\textrm{in}}=\dfrac{i}{\sqrt{2}}(\hat{b}_{j}^{\textrm{in}\dagger}
-\hat{b}_{j}^{\textrm{in}}),
\end{align}
the corresponding linearized QLEs can be obtained in a compact form, i.e.,
\begin{align}
\partial_{t}\hat{u} = A\hat{u}(t)+\hat{v}(t). \label{qfqles}
\end{align}
Here we have introduced the vector of quadrature fluctuation operators $\hat{u}$ and the vector of input noise operators $\hat{v}$, i.e.,
\begin{align}
\hat{u}\!=\!(&\hat{X}_{l}, \hat{Y}_{l}, \hat{X}_{r}, \hat{Y}_{r}, \hat{q}_{l}, \hat{p}_{l}, \hat{q}_{r}, \hat{p}_{r})^{T}, \notag \\
\hat{v}\!=\!(&\hat{X}_{l}^{\textrm{in}}, \hat{Y}_{l}^{\textrm{in}}, \hat{X}_{r}^{\textrm{in}}, \hat{Y}_{r}^{\textrm{in}}, \sqrt{\gamma_{m,l}}\hat{q}_{l}^{\textrm{in}}, \sqrt{\gamma_{m,l}}\hat{p}_{l}^{\textrm{in}}, \sqrt{\gamma_{m,r}}\hat{q}_{r}^{\textrm{in}}, \notag \\
&\sqrt{\gamma_{m,r}}\hat{p}_{r}^{\textrm{in}})^{T},
\end{align}
and the corresponding coefficient matrix
\small{\begin{align}
&A = \notag \\
&\begin{pmatrix}
-\dfrac{\Gamma_{l}}{2} & \tilde{\Delta}_{l,\circlearrowright} & 0 & 0 &
-\Lambda_{l}^{\textrm{im}} & 0 & 0 & 0 \\
-\tilde{\Delta}_{l,\circlearrowright} & -\dfrac{\Gamma_{l}}{2} &
0 & 0 & \Lambda_{l}^{\textrm{re}} & 0 & 0 & 0 \\
-J_{c} & J_{s} & -\dfrac{\Gamma_{r}}{2} &
\tilde{\Delta}_{r,\circlearrowright} & 0 & 0 & -\Lambda_{r}^{\textrm{im}} & 0 \\
-J_{s} & -J_{c} & -\tilde{\Delta}_{r,\circlearrowright} & -\dfrac{\Gamma_{r}}{2} &
0 & 0 & \Lambda_{r}^{\textrm{re}} & 0 \\
0 & 0 & 0 & 0 &
-\dfrac{\gamma_{m,l}}{2} & \omega_{m,l} & 0 & 0 \\
\Lambda_{l}^{\textrm{re}} & \Lambda_{l}^{\textrm{im}} & 0 & 0  & -\omega_{m,l} & -\dfrac{\gamma_{m,l}}{2} & 0 & 0 \\
0 & 0 & 0 & 0 & 0 & 0 &
-\dfrac{\gamma_{m,r}}{2} & \omega_{m,r} \\
0 & 0 & \Lambda_{r}^{\textrm{re}} & \Lambda_{r}^{\textrm{im}} &
0 & 0 & -\omega_{m,r} & -\dfrac{\gamma_{m,r}}{2}
\end{pmatrix},
\end{align}}
with components
\begin{align}
\hat{X}_{j}^{\textrm{in}} = & \sqrt{\kappa_{0,j}}\hat{x}_{0,j}^{\textrm{in}}+\sqrt{\kappa_{\textit{ex},j}}\hat{x}_{\textit{ex},j}^{\textrm{in}},\notag\\
\hat{Y}_{j}^{\textrm{in}} = & \sqrt{\kappa_{0,j}}\hat{y}_{0,j}^{\textrm{in}}+\sqrt{\kappa_{\textit{ex},j}}\hat{y}_{\textit{ex},j}^{\textrm{in}},\notag\\
\Gamma_{j} = & \kappa_{0,j}+\kappa_{\textit{ex},j},\notag\\
\Lambda_{j}^{\textrm{re}} = & 2g_{j}\textrm{Re}[\alpha_{j}],~
\Lambda_{j}^{\textrm{im}} = 2g_{j}\textrm{Im}[\alpha_{j}],\notag\\
J_{s} = & \sqrt{\eta\kappa_{\textit{ex},l}\kappa_{\textit{ex},r}}\sin\varphi,~
J_{c} = \sqrt{\eta\kappa_{\textit{ex},l}\kappa_{\textit{ex},r}}\cos\varphi.
\end{align}
The solution of the linearized QLEs\,(\ref{qfqles}) is given by $\hat{u}(t)=\mathcal{M}(t)\hat{u}(0)+\int_{0}^{t}ds\mathcal{M}(s)\hat{v}(t-s)$, where $\mathcal{M}=\textrm{exp}(At)$. If all real parts of the eigenvalues of $A$ are negative, the system is stable and can reach its steady state, which is determined by the Routh-Hurwitz criterion~\cite{DeJesus1987PRA}. When all the stability condition is satisfied, we can obtain $\mathcal{M}(\infty)=0$ in the steady state and
\begin{align}
\hat{u}_{i}(\infty) = \int_{0}^{\infty}ds\sum_{k}\mathcal{M}_{ik}(s)\hat{v}_{k}(t-s). \label{defu}
\end{align}
Because of the linearized dynamics of the quantum fluctuation quadratures and the Gaussian nature of the input noises, the steady state of this compound system, independently of any initial conditions, can finally evolve into a zero mean quadripartite Gaussian state, which allows us to characterize its steady-state properties by an $8\times8$ covariance matrix (CM) $V$, with matrix elements defined as
\begin{align}
V_{kl} = \left\langle\hat{u}_{k}(\infty)\hat{u}_{l}(\infty)\!+\!\hat{u}_{l}(\infty)\hat{u}_{k}(\infty)\right\rangle/2~(k,l=1,2,\ldots,8).
\end{align}
By using the steady-state solutions in Eq.\,(\ref{defu}) and the fact that the eight components of $\hat{v}(t)$ are uncorrelated with each other, one gets
\begin{align}
V = \int_{0}^{\infty}ds\mathcal{M}(s)D\mathcal{M}^{T}(s), \label{SSCMV}
\end{align}
where $D=(1/2)\,\textrm{Diag}[\Gamma_{l},\Gamma_{l},\Gamma_{r},\Gamma_{r},\gamma_{m,l}(2\bar{n}_{m,l}+1)
,\gamma_{m,l}(2\bar{n}_{m,l}\\ +1),\gamma_{m,r}(2\bar{n}_{m,r}+1),\gamma_{m,r}(2\bar{n}_{m,r}+1)]$ is the diffusion matrix, and it is defined through $D_{kl}\delta(s-s^{\prime})=\langle\hat{v}_{k}(s)\hat{v}_{l}(s^{\prime})+\hat{v}_{l}(s^{\prime})\hat{v}_{k}(s)\rangle/2$. When the stability condition is fulfilled, the steady-state CM $V$ in Eq.\,(\ref{SSCMV}) is determined by the Lyapunov equation
\begin{align}
AV + VA^{T} = -D. \label{LE}
\end{align}
Clearly, it is seen that the Lyapunov equation\,(\ref{LE}) is linear for $V$ and can be solved straightforwardly, thereby implying that one can derive the CM $V$ for any specific value of relevant parameters.

\begin{figure*}[htbp]
\centering
% Requires \usepackage{graphicx}
\includegraphics[width=1\textwidth]{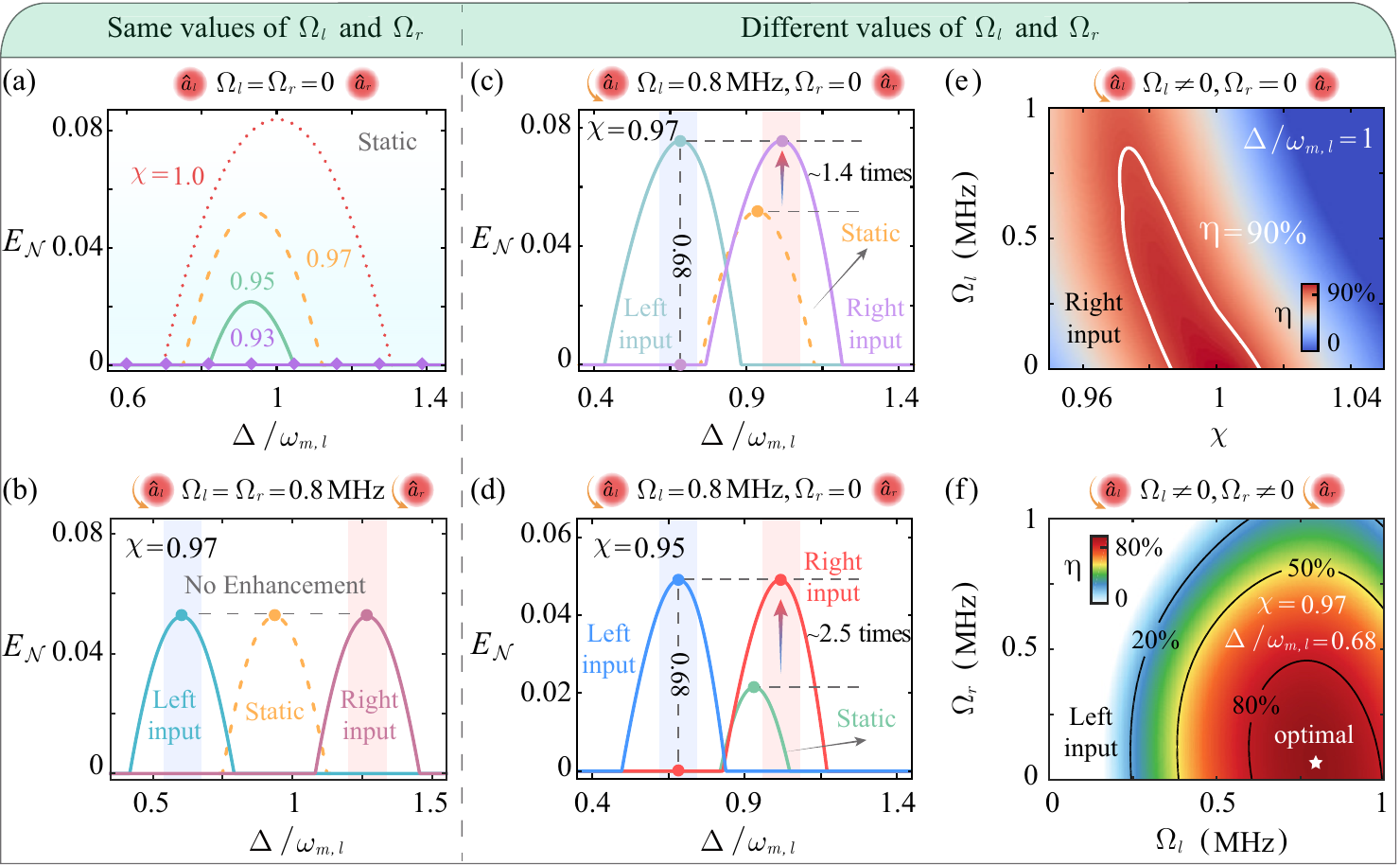}
\caption{\label{Fig_enhancement}Suppression of remote entanglement between two nonidentical mechanical oscillators due to their frequency mismatch, and its nonreciprocal revival resulting from the rotation-induced compensation. The logarithmic negativity $E_{\mathcal{N}}$ is plotted as a function of the scaled optical detuning $\Delta/\omega_{m,l}$ for different input directions, with angular velocities of the left- and right-hand-side WGM resonators (a) $\Omega_{l}=\Omega_{r}=0$, (b) $\Omega_{l}=\Omega_{r}=0.8\,\textrm{MHz}$, and (c)-(d) $\Omega_{l}=0.8\,\textrm{MHz},~\Omega_{r}=0$. For static cascaded systems, $E_{\mathcal{N}}$ tends to be inhibited as the mechanical frequency ratio $\chi$ increases. In comparison, by spinning one of the WGM resonators, the maximum value of $E_{\mathcal{N}}$ can be enhanced by about $1.4$ or $2.5$ times with respect to $\chi=0.97$ or $0.95$, almost reaching the values of the case with identical mechanical frequencies. (e) Density plot of the revival factor $\eta$ as a function of the mechanical frequency ratio $\chi$ and the angular velocity $\Omega_{l}$, with light input from the right-hand side, $\Omega_{r}=0$ and $\Delta/\omega_{m,l}=1$. (f) Density plot of the revival factor $\eta$ as a function of the angular velocities $\Omega_{l}$ and $\Omega_{r}$, with light input from the left-hand side, $\eta=0.97$, and $\Delta/\omega_{m,l}=0.68$.}
\end{figure*}

\section{Nonreciprocal remote entanglement between two mechanical oscillators}\label{SecIII}
In order to quantify the bipartite entanglement of a multimode continuous variable (CV) system, one can adopt the logarithmic negativity $E_{\mathcal{N}}$ as an effective measure, which is defined as~\cite{Adesso2004pra}
\begin{align}
E_{\mathcal{N}}=\max\,[0,-\ln (2\nu^{-})], \label{eq15}
\end{align}
where $\nu^{-}\!=\!2^{-1/2}\{\Sigma(V_{\mu\nu})-[\Sigma(V_{\mu\nu})^{2}-4\det\!V_{\mu\nu}]^{1/2}\}^{1/2}$ [with $\Sigma(V_{\mu\nu})\!\equiv\!\det\mathcal{A}+\det\mathcal{B}-2\det\mathcal{C}$] is the minimum symplectic eigenvalue of the partial transpose of a reduced $4\times4$ CM, $V_{\mu\nu}$, with $\mu$ and $\nu $ describing the two selected modes under consideration. The reduced CM $V_{\mu\nu}$ contains the entries of $V$ associated with the corresponding bipartition $\mu$ and $\nu$, which can be obtained straightforwardly by tracing out the rows and columns of the uninteresting modes in $V$. By writing the reduced CM $V_{\mu\nu}$ in a $ 2\times2 $ block form, we have
\begin{align}
V_{\mu\nu}=\left(
\begin{matrix}
\mathcal{A}&\mathcal{C}\\
\mathcal{C}^{\textit{T}}&\mathcal{B}
\end{matrix}
\right).\label{reducedCM}
\end{align}
Note that the physicality of the reduced CM $V_{\mu\nu}$ can be thoroughly checked by considering the fulfillment of the Heisenberg-Robinson uncertainty principle, that is, checking whether the minimum symplectic eigenvalue $\nu^{-}$ satisfies the condition that $|\nu^{-}|\geq1/2$. In terms of the case with $\nu^{-}<1/2$, the corresponding bipartition of the considered system gets entangled, which is equivalent to Simon's necessary and sufficient entanglement nonpositive partial transpose criterion (or the related Peres-Horodecki criterion) for certifying bipartite entanglement in Gaussian states~\cite{Simon2000PRL}. Correspondingly, the logarithmic negativity, $E_{\mathcal{N}}$, quantifying the amount by which the Peres-Horodecki criterion is violated, is an efficient measure that is widely used for certifying bipartite entanglement in CV systems.

In the following discussions, we first demonstrate how to achieve nonreciprocal remote entanglement between two identical mechanical oscillators (i.e., $\delta\omega_{m}=0$ or $\chi=1$). Specifically, as shown in Fig.\,\ref{Fig_nonreciprocity}, the logarithmic negativity $E_{\mathcal{N}}$ for opposite input directions is plotted as a function of the scaled optical detuning $\Delta/\omega_{m,l}$ with respect to different angular velocities of the two WGM resonators. In our numerical simulations, for ensuring the stability of this compound system, we assume that the following experimentally feasible parameters are employed for two identical WGM resonators~\cite{Maayani2018Nature,Shen2016NP}: $n=1.48$, $m=15\,\textrm{ng}$, $R=36\,\mu\textrm{m}$, $\lambda=780\,\textrm{nm}$, $\kappa_{0}/2\pi=15\,\textrm{MHz}$, $\omega_{m}/2\pi=88.54\,\textrm{MHz}$, $\gamma_{m}/2\pi=2.2\,\textrm{kHz}$, $P=20\,\textrm{mW}$, $T=100\,\textrm{mK}$, $\varphi=0$, $\eta=1$, and $\Omega = 0.6\,\textrm{MHz}$. Also, the fiber-cavity coupling rates for the left- and right-hand-side WGM resonators are chosen as $\kappa_{\textit{ex},l}/2\pi=27\,\textrm{MHz}$ and $\kappa_{\textit{ex},r}/2\pi=30\,\textrm{MHz}$, respectively. Moreover, due to the fact that the radiation-pressure-induced frequency shift is much smaller than the cavity-pump detuning, i.e., $g_{0,j}(\beta_{j}^{\ast}+\beta_{j})\ll|\omega_{c}-\omega_{d}|$, we have already set $\Delta_{c,l}=\Delta_{c,r}=\Delta$ for simplicity.

In the case of static cascaded systems (i.e., $\Omega_{l}=\Omega_{r}=0$), as shown in Fig.\,\ref{Fig_nonreciprocity}(b), $E_{\mathcal{N}}$ is present only within a small interval of $\Delta$ around the COM resonance $\Delta/\omega_{m,l}=1$. Besides, it is also seen that $E_{\mathcal{N}}$ stays reversible when switching the driving directions under the same parameter conditions [see the solid and circle lines in Fig.\,\ref{Fig_nonreciprocity}(b)], which implies that the generation of the entanglement between mechanical oscillators is reciprocal for static cascaded systems. In contrast, as shown in Fig.\,\ref{Fig_nonreciprocity}(c), by spinning one arbitrary WGM resonator in this compound system (e.g., $\Omega_{l}\neq0$ and $\Omega_{r}=0$), $E_{\mathcal{N}}$ becomes irreversible with regards to light input from opposite driving directions, indicating the achievement of nonreciprocal entanglement of the associated mechanical oscillators. This result implies that when a driving field with a specific frequency is applied to this compound system, one can entangle the two mechanical oscillators through the radiation-pressure-induced interactions with respect to one specific input direction, whereas no entanglement occurs under the same parameter conditions for an opposite input direction [see $E_{\mathcal{N}}$ for different driving directions at e.g., $\Delta/\omega_{m,l}=0.74$ or $0.98$ in Fig.\,\ref{Fig_nonreciprocity}(c)]. However, it is also seen that the degree of such nonreciprocal entanglement is suppressed in comparison with the static situation. In addition, as shown in Fig.\,\ref{Fig_nonreciprocity}(d), it is found that in the case of two WGM resonators spinning along the same direction, $E_{\mathcal{N}}$ can also be irreversible for opposite driving directions, which provides an alternative way to achieve such nonreciprocal entanglement. Besides, compared with the former case of a single spinning resonator, the generated entanglement in this situation acquires a higher entanglement degree, which is the same as that of a static system. In regards to two oppositely spinning WGM resonators, we have also confirmed that the generation of such entanglement is reversible due to the unbroken time-reversal symmetry, and thus these results will not be reported here. The underlying physics for the generation of nonreciprocal entanglement can be understood as follows. In static cascaded systems, light experiences the same travel time and shares the same resonant wavelength with respect to the left and right input directions, which is guaranteed by the time-reversal symmetry of this compound system. Correspondingly, the optical-radiation-pressure-mediated generation of the entanglement between mechanical oscillators is reciprocal in this regime. However, by selectively spinning the WGM resonators and properly adjusting their spinning directions, the time-reversal symmetry of this compound system is thus broken due to the presence of the Sagnac effect. In this situation, the resonant wavelength becomes irreversible for opposite driving directions, which further results in the direction-dependent COM interactions for each WGM resonator. Therefore, for driving light with a specific wavelength, it can be observed that the generation of such optical-radiation-pressure-mediated entanglement is nonreciprocal.

\begin{figure}[htbp]
\centering
% Requires \usepackage{graphicx}
\includegraphics[width=0.47\textwidth]{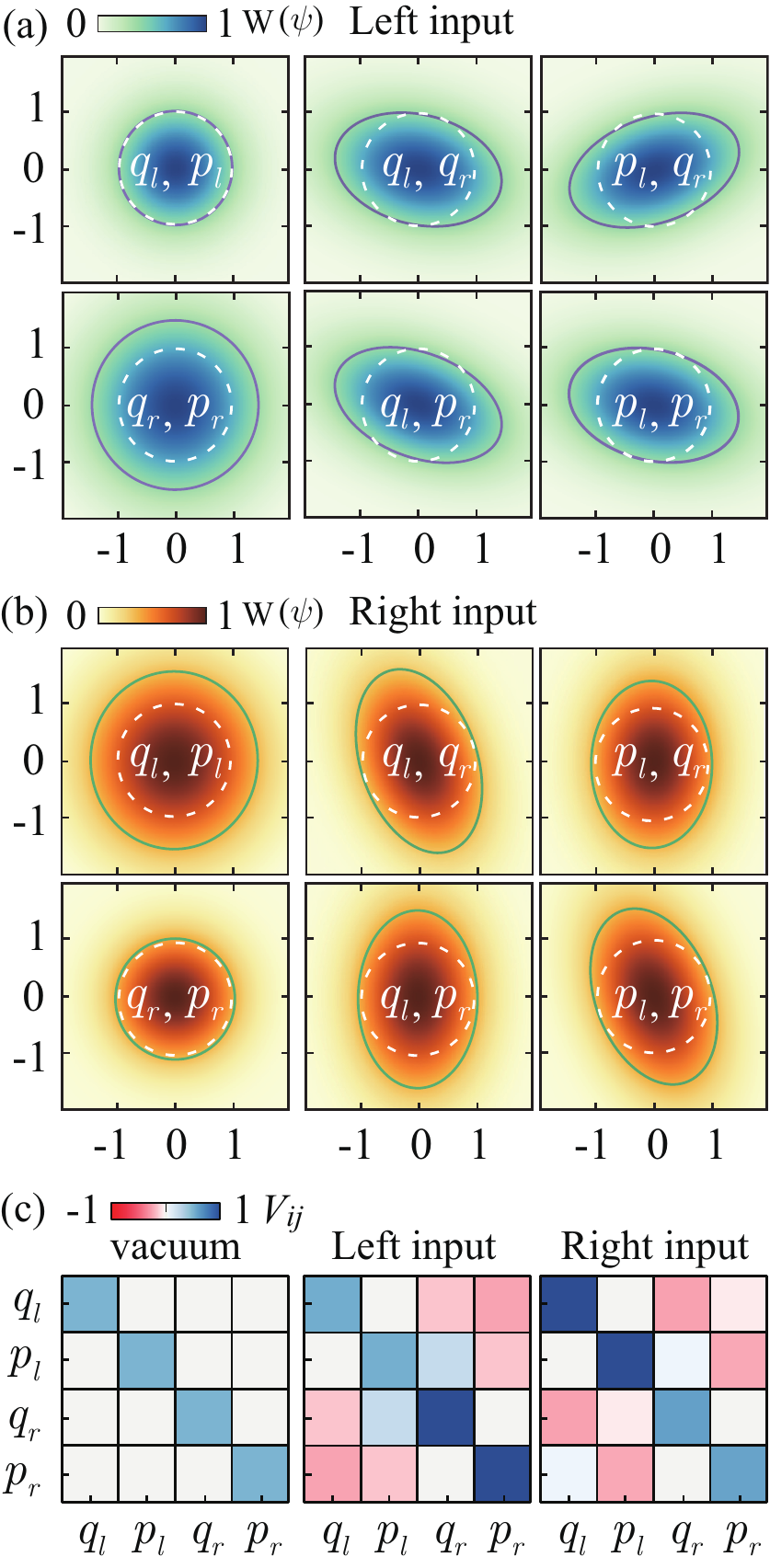}
\caption{\label{Fig_stastics}Nonreciprocal quadrature statistics of two nonidentical mechanical oscillators in cascaded spinning COM systems. (a), (b) Reconstructed Wigner function $\textrm{W}(\psi)$ for the steady state of the associated mechanical oscillators, with respect to the case where the driving field is input from the left- and right-hand sides, respectively, along with that of an ideal vacuum state for reference. Each panel corresponds to a characteristic projection of $\textrm{W}(\psi)$ on two-dimensional subspaces within $\psi \!=\! (q_{l}, p_{l}, q_{r}, p_{r})^{T}$, which is obtained by integrating out the other two quadratures in $\psi$. The solid (dashed) line indicates a drop of $1/\textrm{e}$ in the maximum value of $\textrm{W}(\psi)$ for the relevant steady (vacuum) state. The legends $\psi_{i}$, $\psi_{j}$ in each panel correspond to a correlation with $\psi_{i}$ along the $\textrm{x}$ axis and $\psi_{j}$ along the $\textrm{y}$ axis. (c) Reduced CM $V_{\mu\nu}$ of the two mechanical oscillators, which is associated with the CM used in (a) and (b). We set the optical detuning $\Delta/\omega_{m,l}=0.68$ and the mechanical frequency ratio $\chi=0.95$. The other parameters are the same as in the text.}
\end{figure}

In Fig.\,\ref{Fig_enhancement}, we further show the unique properties of nonreciprocal entanglement by which one can enhance the entanglement degree between two frequency-mismatched mechanical oscillators, thereby providing an enticing opportunity to efficiently generate, manipulate and exploit entangled states of nonidentical mechanical objects at a wide range of frequencies. Figure \ref{Fig_enhancement}(a) shows that, in the case of a static cascaded system driven by a single-tone laser, $E_{\mathcal{N}}$ achieves its maximum value just at the mechanical frequency resonance, i.e., $\chi=1$, which is consistent with the results shown in Refs.~\cite{Mancini2003EPJD,Genes2008NJP}. This result means that such mechanical entanglement tends to be inhibited or even vanish when one increases the frequency mismatch between the associated mechanical oscillators. For recent theoretical and experimental studies~\cite{Li2015NJP,Lai2022PRL,Riedinger2018Nature,Korppi2018Nature,Kotler2021Science,Mercier2021Science}, an effective approach for entangling two frequency-mismatched mechanical oscillators is to apply a multitone driving laser, where each tone of the driving laser is associated with a specific mechanical oscillator to generate correlations between them. However, the multitone driving scheme is generally restrained by the RWA approximation~\cite{Li2015NJP}, which requires the fundamental mechanical frequencies $\omega_{m}$ and the corresponding frequency difference $\delta\omega_{m}$ to be much larger than the cavity bandwidth $\kappa$, i.e., $\omega_{m},\,\delta\omega_{m}\gg\kappa$. As a result, to fulfill this parameter condition of the RWA approximation, it usually requires ultrahigh optical quality factors and a large frequency difference for the associated mechanical oscillators. However, based on optical-wavelength microcavities, it still remains challenging for current experimental techniques to fabricate the COM devices with such characteristics. Here we further study how to entangle two such nonidentical mechanical oscillators by nonreciprocal control. As shown in Fig.\,\ref{Fig_enhancement}(b), by spinning the two WGM resonators along the same direction, it is seen that although nonreciprocal entanglement can also be generated in this situation, the entanglement degree is not enhanced. However, as shown in Figs.\,\ref{Fig_enhancement}(c) and \ref{Fig_enhancement}(d), by spinning one arbitrary WGM resonator (e.g., $\Omega_{l}\neq0$ and $\Omega_{r}=0$), it is found that the maximum values of $E_{\mathcal{N}}$ are considerably improved for one specific input direction in this cascaded spinning system, surpassing approximately a $1.4$ or $2.5$ times enhancement in the degree of entanglement compared to that of the static case, while for the opposite input direction, $E_{\mathcal{N}}$ is correspondingly inhibited. These results can be understood as follows. For static cascaded systems, due to the frequency mismatch of the two mechanical oscillators, the COM resonance condition is fulfilled merely by one specific WGM resonator with respect to the single-tone driving laser. Consequently, the effective interaction between the two mechanical oscillators is suppressed, which further leads to an inhibition of the optical-radiation-pressure-mediated entanglement generation. In contrast, by means of selectively spinning the WGM resonators, the corresponding COM resonance condition is separately optimized for each WGM resonator, which provides facilities to the enhancement of such entanglement. In order to further quantify the amount by which the entanglement degree is enhanced, we define the following revival coefficient:
\begin{align}
\eta=\dfrac{\max[E_{\mathcal{N}}(\Omega\neq0,\delta\omega\neq0)]}{\max[E_{\mathcal{N}}(\Omega=0,\delta\omega=0)]}.
\end{align}
As clearly seen in Figs.\,\ref{Fig_enhancement}(e) and \ref{Fig_enhancement}(f), the revival coefficient $\eta$ can even reach values above $80\%$ or $90\%$ for specific frequency mismatches, which is otherwise attainable in static cascaded systems. This result indicates that by carefully adjusting the angular velocities of such spinning devices, one can create nearly ideal remote entanglement between nonidentical mechanical oscillators. More importantly, it is stressed that our numerical calculations of $E_{\mathcal{N}}$ with respect to frequency-mismatched oscillators are performed without any parameter constraints required by the RWA approximation, which can facilitate the achievement of such entanglement between nonidentical mechanical oscillators within current optical-wavelength COM devices.

After having established nonreciprocal entanglement, we now investigate the associated nonreciprocal quadrature statistics of the two relevant nonidentical mechanical oscillators. For this purpose, we plot a selection of characteristic projections of the reconstructed Wigner function $\textrm{W}(\psi)$ and their corresponding CM $V_{\mu\nu}$ in Fig.\,\ref{Fig_stastics}. Here, as a specific example, we have taken $\Delta/\omega_{m,l}=0.68$ and $\chi=0.95$, while the other parameters are the same as in Fig.\,\ref{Fig_enhancement}. Because the steady state of this compound system belongs to the class of Gaussian states, its reconstructed Wigner function $\textrm{W}(\psi)$ can thus be characterized by a multivariate normal distribution, which is defined as
\begin{align}
\textrm{W}(\psi)=\dfrac{\exp\left[-\dfrac{1}{2}\left(\psi^{\dagger}V_{\mu\nu}^{-1}\psi\right)\right]}{\pi^{2}\sqrt{\det(V_{\mu\nu})}}.
\end{align}
Figures \ref{Fig_stastics}(a) and \ref{Fig_stastics}(b) demonstrate the characteristic projections of $\textrm{W}(\psi)$ on two-dimensional subspaces with respect to the left and right input directions, respectively, along with that of an ideal vacuum state for reference. The ellipse (circle) with solid (dashed) line indicates a drop of $1/\textrm{e}$ in the maximum value of $\textrm{W}(\psi)$ for the relevant steady (vacuum) state of this compound system. When the boundary of the ellipse enters the circle, it indicates a situation where the variance of the relevant state is smaller than that of a vacuum state, and thus such a state is a squeezed state. In the case of the left input direction, it is found that the reconstructed Wigner function $\textrm{W}(\psi)$ exhibits all typical features of an entangled state: (i) the projections of $\textrm{W}(\psi)$ for local quadrature pairs of each single mechanical oscillator is consistent with a Gaussian distribution of large variance, resulting from the phase-preserving amplification of the vacuum state [see the left column of Fig.\,\ref{Fig_stastics}(a)]; (ii) the projection of $\textrm{W}(\psi)$ for cross-quadrature pairs, e.g., $\{p_{l},q_{r}\}$ or $\{q_{l},p_{r}\}$, is squeezed along the diagonal direction, indicating the emergence of two-mode squeezing correlations between the associated mechanical oscillators [see the middle and right columns of Fig.\,\ref{Fig_stastics}(a)]. In contrast, for the right input direction, it is found that none of the characteristic projections of $\textrm{W}(\psi)$ show squeezing correlation for quadrature pairs [see Fig.\,\ref{Fig_stastics}(b)]. These results are consistent with the features of the nonreciprocal entanglement previously shown in Fig.\,\ref{Fig_enhancement}. Besides, as seen in Fig.\,\ref{Fig_stastics}(c), the corresponding CMs $V_{\mu\nu}$ with respect to the left and right input directions are both characterized by four dominant diagonal elements and eight minor off-diagonal elements, with $V_{13}\approx V_{24}$, $V_{14}\approx -V_{23}$, $V_{31}\approx V_{42}$, and $V_{32}\approx -V_{41}$, which clearly display the correlations and anticorrelations between the cross-quadrature pairs. Also, from these results, one finds that the correlations of cross-quadrature pairs for the case of the right input direction are smaller than those of its left input counterparts, which is one of the possible reasons for the achievement of nonreciprocal entanglement in the previous discussions.

Finally, we remark that by utilizing a single spinning COM resonator, we have proposed how to achieve nonreciprocal entanglement between optical field and mechanical motion very recently, and revealed its counterintuitive property as being robust against random losses~\cite{Jiao2020PRL}. In this paper, we mainly focus on the generation and manipulation of nonreciprocal entanglement between two massive mechanical oscillators and investigate how to enhance the degree of such entanglement by exploiting the nonreciprocal control of a cascaded system.

\section{Conclusion}\label{SecIV}

In summary, we present an effective method to generate and manipulate nonreciprocal entanglement between two spatially separated mechanical oscillators in a cascaded COM system, and reveal its unique properties by which one can entangle two frequency-mismatched mechanical oscillators through a single-tone driving laser with considerable entanglement degrees, which are otherwise unattainable in its static counterparts. The Sagnac effect induced by spinning devices play a key role in our proposed scheme. It allow us to break the time-reversal symmetry of this compound system, and to also optimize the optical-radiation-pressure-mediated interactions between mechanical oscillators. As a result, by selectively spinning the WGM resonators, we observ the achievement of nonreciprocal entanglement, and more excitingly, the enhancement of such entanglement between two nonidentical mechanical oscillators. Besides, we also find that the quadrature statics of the associated mechanical oscillators is simultaneously nonreciprocal as well, that is, the two-mode squeezing correlations for cross-quadrature pairs is present merely with regards to one specific input direction. These findings, combining the concepts of nonreciprocal physics~\cite{Sounas2017NP,Lodahl2017Nature,Caloz2018PRApp} and quantum states engineering~\cite{Horodecki2009RMP,Andersen2010LPR,Nielsen2010book,Barzanjeh2022NP,Flamini2018RPP}, provide an enticing opportunity to generate, manipulate and harness the macroscopic entanglement between distinct massive mechanical objects~\cite{Mancini2002PRL,Mancini2003EPJD,Genes2008NJP,Huang2009NJP,Tan2013PRA,Liao2014PRA,Yang2015PRA,
Li2015NJP,Lai2022PRL,Riedinger2018Nature,Korppi2018Nature,Kotler2021Science,Mercier2021Science}. On experimental aspect, the realization of our proposed scheme involves the process of breaking time-reversal symmetry using the Sagnac effect, which is based on spinning optical devices or moving medium~\cite{Maayani2018Nature,Lai2019Nature,Mao2022arxiv}. We note that the exploration of such nonreciprocal entanglement can also be extended to acoustic~\cite{Fleury2014Science,Wang2022SA}, plasmonic~\cite{Dong2021Nature} or thermal systems~\cite{Xu2020NC} by using the analog of the Sagnac effect. Furthermore, we also envision that, instead of using spinning devices as described here, future developments with the engineering of nonreciprocal quantum entanglement can also be implemented based on varieties of other nonreciprocal devices that involve atomic ensembles~\cite{Zhang2018NP,Yang2019PRL,Liang2020PRL}, nonlinear optical cavities~\cite{Dong2015NC,Kim2015NP,Xia2018prl,Hamann2018PRL,TangL2022PRL}, COM systems~\cite{Manipatruni2009PRL,Shen2016NP,Bernier2017NC,Mercier2019prapp}, synthetic structures~\cite{Peng2014NP,Chang2014NP,Xu2020PRApp,Chen2021PRL}, and hot propagating modes~\cite{Orr2022arxiv}. In a broader view, although we have considered here a specific case of single-tone optical driving, we stress that our proposed scheme can be further combined with e.g., multitone lasers, to explore a more efficient and versatile way for the generations and manipulations of nonreciprocal entanglement~\cite{Riedinger2018Nature,Korppi2018Nature,Kotler2021Science,Mercier2021Science}. As such, we believe that our proposed scheme is poised to serve as a useful tool for conceptual exploration of quantum fundamental theories~\cite{Horodecki2009RMP}, facilitate the manipulation of various kinds of couplings between remote quantum systems within cascaded configurations~\cite{Xiao2010PRA,Li2016OE,Santos2017PRL,Zhang2015PRL,Li2022PRL}, and provide appealing quantum resources for emerging quantum technologies ranging from quantum information processing~\cite{Andersen2010LPR,Nielsen2010book,Flamini2018RPP} to quantum sensing~\cite{Degen2017RMP}.

\section{Acknowledgment}

The authors thank Jie-Qiao Liao, Xun-Wei Xu and Yun-Lan Zuo for helpful discussions. H. J. is supported by the National Natural Science Foundation of China (NSFC, Grants No. 11935006 and 11774086) and the Science and Technology Innovation Program of Hunan Province (Grant No. 2020RC4047). L.-M. K. is supported by the NSFC (Grants No. 1217050862, 11935006 and 11775075). Y.-F. J. is supported by the NSFC (Grant No. 12147156), the China Postdoctoral Science Foundation (Grants No. 2021M701176 and 2022T150208) and the Science and Technology Innovation Program of Hunan Province (Grant No. 2021RC2078).

%\bibliography{myreference.bib}

%

\end{document}